\documentclass[twocolumn,prb,superscriptaddress,floatfix]{revtex4}
\usepackage{color}
\usepackage{colordvi}
\usepackage{amsmath}
\usepackage{amssymb}
\usepackage{fancybox}
\usepackage{epsfig}
\usepackage{graphicx}
\usepackage{tabularx}
\begin{document}

\title{The role of disorder symmetry in a dirty superconductor: a
Bogoliubov-deGennes (BdG) study}
\author{Sanjeev Kumar}
\email{sanjeev@iisermohali.ac.in}
\affiliation{Department of Physical Sciences, Indian Institute of Science Education
and Research (Mohali), 
Knowledge city, Sector 81, SAS Nagar, Mohali 140306, India} 
\author{Prabuddha B. Chakraborty}
\affiliation{ Indian Statistical Institute, Chennai Centre, SETS Campus, MGR
Knowledge City,
Taramani, Chennai 600113, India} 

\begin{abstract}
We study the effect of disorder symmetry in a disordered s-wave superconductor. We
begin with an attractive Hubbard model on a two-dimensional square lattice in the
presence of diagonal- and off-diagonal disorder. The model is studied numerically
using Bogoliubov-deGennes approach, which proceeds via the decoupling of attractive Hubbard 
term in the pairing channel. We find that the off-diagonal disorder is much
more effecient in destroying superconducting order. A detailed analysis brings out
very distinct qualitative pictures in the two cases: diagonal disorder leads to
formation of small scale isolated superconducting islands separated by large region
of normal metal, whereas off-diagonal disorder leads to a connected network of
superconducting region. Naively, it seems that the connected network of
superconducting regions should have a large average value for superconducting order
parameter, but this does not hold true for large disorder, as we demonstrate
explicitly. Our qualititive picture is supported by real-space data on local
order parameters.

\end{abstract}
\date{\today}

\maketitle

\underline{Introduction:} The behavior of a fermion system in the presence of
disorder and strong inter-particle interactions has been a challenging question in
condensed matter physics for many decades.\cite{Lee85,Altshuler85,Belitz94}
Anderson's seminal discovery that the wave nature of electrons in a quantum
mechanical context can cause the electrons to completely localize in a disordered
system gave rise to the field of Anderson
localization.\cite{Anderson58,Anderson50-2010} On the other hand, in the clean
system, the Coulomb repulsion between the electrons themselves can also cause the
electrons to localize, a phenomenon known as Mott localization,\cite{Mott90} the
most famous example of such materials being the undoped parent compounds of the
high-T$_C$ Cuprate superconductors.\cite{Lee2006}

A well-studied problem in condensed matter physics has been the properties of a
dirty superconductor, with early theories by Anderson\cite{Anderson:1959} and
Abrikosov and Gorkov\cite{Abrikosov:1959} in the regime when the disorder was weak.
With the advent of advanced experimental techniques of growing highly inhomogenoeous
films, research on the properties of a dirty superconductor in the strong disorder
regime has become possible.\cite{Hebard:1994,Goldman:1998} This is also the regime
where electron interactions start to play an important role, which then becomes a
very challenging theoretical problem.\cite{Belitz94} Also, controlled studies of the
related quantum phase transition between the superconductor and the Anderson
Insulator (SIT) have been widely carried out. In the SIT, the early theoretical
prediction of a universal conductivity (in units of
$e^{2}/\hbar$)\cite{Grinstein:1990} demarcating the critical point between the two
phases generated a plethora of theoretical and experimental work in this
field.\cite{Minchul:1991,Sorensen:1992,Wallin:1994,Bollinger:2011} Recently, with
the observation of Anderson localization in disordered optical lattices and the
possibility of tuning the sign of the interaction between the atomic species, a new
avenue has opened up to investigate the interplay of disorder and interaction among
both fermionic and bosonic
atoms.\cite{Bloch:2008,Aspect:2009,Sanchez-Palencia:2010,Jendrzejewski:2012} 

In the presence of attractive interactions between fermions, disordered systems can
become superconducting (for charged electrons) or superfluid (for neutral atoms).
Below a critical disorder strength in $d=2$, the two-particle states may sustain
dissipationless flow even though all single-particle states are localized due to
Anderson localization. The combination of attractive interactions and disorder can
lead to a rich variety of phases and new physics, including the BCS-BEC crossover,
various kinds of glassy phases and superfluids (superconductors). 

A related very important aspect of physical systems with disorder is the symmetry of
the Hamiltonian governing them. A complete classification of the symmetries of the
single-particle Hamiltonians governing a disordered system was recently
achieved.\cite{Zirnbauer:1996,Altland:1997,Evers:2008} Depending on the presence (or
absence) of time-reversal symmetry ($T$), spin-rotation ($SU(2)$) symmetry and
particle-hole symmetry ($\Xi$) in the Hamiltonian, the physics of the disordered
system can be markedly different, particularly with respect to Anderson
localization. For example, it is now known that weak localization does not occur in
the chiral symmetry class in the conventional sense, but follows a topological route
which then gives way to strong localization for stronger disorder.\cite{Konig:2012}
 
In traditional solid state systems, intrinsic disorder typically originates from
random locations of dopant ions. This invariably affects both the on-site potential
and the hopping parameters in the corresponding model Hamiltonian. Depending on the
details of the structure and ions involved, this may affect one or the other more
strongly and therefore can be modelled via either a randomness in on-site potentials
(diagonal disorder) or hopping strengths (off-diagonal disorder). In addition,
disordered optical lattices provide another promising direction to investigate the
influence of symmetries on the localization problem.  

The SIT has been investigated in considerable detail in the literature with
orthogonal disorder i.e., on-site or diagonal disorder with and without the
spin-rotation (SU(2))
symmetry.\cite{Ghosal:1998,Ghosal:2001,Bouadim:2011,Nanguneri:2012} In this paper,
we investigate the influence of off-diagonal disorder in the physics of a dirty
superconductor, which belongs, in the classification of Altland and
Zirnbauer\cite{Altland:1997} in the chiral-BdG symmetry class, and compare it to
similar physics in the usually studied BdG-orthogonal symmetry class. Even though we
do not explicitly investigate the physics of the critical point itself, the effect
of imminent localization is always present in the physics of a dirty superconductor,
and the nature and the properties of the inhomogeneities in the superconducting
state itself is a very important indicator of the subsequent physics of
localization. Elucidating how the superconducting state responds to the nature of
the disorder potential will be the main goal of this paper.                     

\underline{Model and Method:} The starting point of our investigation is the
attractive Anderson-Hubbard model on a square lattice, given by the Hamiltonian
\begin{equation}
H = H_{\rm{sp}} + H_{\rm{int}},
\label{Hamiltonian0}
\end{equation}
where $H_{\rm{sp}}$ is the single-particle part and $H_{\rm{int}}$ is the
interaction part. $H_{\rm{sp}}$ is given by
\begin{equation}
H_{\rm{sp}} = -\sum_{\left<ij\right>,\sigma}t_{ij}c_{i\sigma}^\dagger c_{j\sigma} -
\mu\sum_{i\sigma}c_{i\sigma}^\dagger c_{i\sigma} +
\sum_{i\sigma}\varepsilon_{i}c_{i\sigma}^\dagger c_{i\sigma} .
\label{Hamiltonian-sp}
\end{equation}
Here $c_{i\sigma}\left(c_{i\sigma}^\dagger\right)$ annihilates (creates) an electron
at site $\vec{R}_{i}$ with spin-projection $\sigma$, $\left<ij\right>$ implies that
$\vec{R}_{i}$ and $\vec{R}_{j}$ are nearest neighbors, $\mu$ is the chemical
potential and $t_{ij}$ is the hopping matrix element between sites $\vec{R}_{i}$ and
$\vec{R}_{j}$. We write $t_{ij}\left(=t_{ji}\right)$ as 
\begin{equation}
t_{ij} = t + \delta_{ij},
\label{randomness}
\end{equation}
where $\delta_{ij}$ is drawn randomly from a box probability distribution
$P\left(\delta_{ij}\right) = \frac{1}{V_{t}}\Theta\left(\frac{V_{t}}{2} -
|\delta_{ij}|\right)$, $\Theta\left(x\right)$ being the Heaviside $\Theta$-function.
Similarly, the on-site diagonal disorder term is denoted by the random site-energy
$\varepsilon_{i}$, which is drawn from a box-distribution given by
$P\left(\varepsilon_{i}\right) = \frac{1}{V}\Theta\left(\frac{V}{2} -
|\varepsilon_{i}|\right)$. The attractive Hubbard interaction term $H_{\rm{int}}$ is
given by
\begin{equation}
H_{\rm{int}} = -\left|U\right|\sum_{i} n_{i\uparrow}n_{i\downarrow},
\label{Hamiltonian-int}
\end{equation}
where $n_{i\sigma}=c_{i\sigma}^{\dagger}c_{i\sigma}$ is the number operator at site
$\vec{R}_{i}$ and spin-projection $\sigma$. We set $t=1$ as the basic scale for energy, therefore
all energy parameters are in units of $t$.
This leaves four independent energy-scales in the model: the interaction strength
$\left|U\right|$, the off-diagonal disorder bandwidth $V_{t}$ , the diagonal
disorder bandwidth $V$ and the temperature $T$. If $V_{t}=0$, the model reduces to
the class of disorder Hamiltonians with orthogonal symmetry. If $V=0$, it reduces to
the class of disorder Hamiltonians with chiral symmetry.

At half-filling, the clean (i.e., no disorder) attractive Hubbard model (in $d=2$)
has a phase diagram in which the charge density wave (CDW) state and the
superconducting state (SC) are exactly degenerate in energy and co-exist at $T=0$.
As one dopes the system away from half-filling, the CDW state gradually disappears,
and the system makes a Berezinskii-Kosterlitz-Thouless (BKT) transition to a
conventional s-wave SC ground state.\cite{Robaszkiewicz:1981,Scalettar:1989}

A discussion of the effect of non-magnetic disorder on superconductivity (in
particular, $T_{C}$) begins with Anderson's theorem,\cite{Anderson:1959} which
states that, at least for weak impurity scattering, $T_{C}$ is not affected by
disorder. In the context of  mean-field BdG formalism with diagonal disorder, Ghosal
{\textit{et al}}\cite{Ghosal:1998,Ghosal:2001} found that it is impossible to
destroy superconductivity through amplitude fluctuations by increasing disorder (an
extreme case of Anderson's theorem), and in fact, there is a disorder strength for
which the spectral gap (the superconducting gap) shows a non-zero minimum, beyond
which the gap actually increases with disorder strength, even though the spatially
averaged SC order parameter monotonically vanishes. Conventional wisdom suggests
that the impurity states will gradually fill in the superconducting gap and destroy
SC. However, as we shall see, this intuitive picture is not always true, and in
fact, depends on the type of disorder.

We revisit this scenario where the disorder resides on the bonds
(off-diagonal/bond/hopping disorder) rather than on the sites
(diagonal/site/potential disorder). The BdG equations are derived as usual, by
decoupling the attractive Hubbard interaction in the pairing channel. This leads to
the effective BdG Hamiltonian:\cite{deGennes:1963}
\begin{eqnarray}
H_{\rm{eff}} & = & -\sum_{\left<ij\right>,\sigma}t_{ij}c_{i\sigma}^\dagger
c_{j\sigma} - \sum_{i\sigma}\tilde{\mu}_{i\sigma}n_{i\sigma} +
\sum_{i\sigma}\varepsilon_{i}n_{i\sigma}\nonumber \\
& & + \sum_{i}\left[\Delta_{i}c_{i\uparrow}^\dagger c_{i\downarrow}^\dagger +
\Delta^{\ast}_{i}c_{i\downarrow} c_{i\uparrow}\right],
\label{BdGequation}
\end{eqnarray}        
where $\tilde{\mu}_{i\sigma} = \mu + \left|U\right|\left<n_{i,-\sigma}\right>$.
$H_{\rm{eff}}$ is diagonalized via the Bogoliubov transformation:
\begin{eqnarray}
c_{i\uparrow} = \sum_{n}\left[\gamma_{n\uparrow}u_{in} -
\gamma_{n\downarrow}^{\dagger}v^{\ast}_{in}\right]\nonumber\\
c_{i\downarrow} = \sum_{n}\left[\gamma_{n\downarrow}u_{in} +
\gamma_{n\uparrow}^{\dagger}v^{\ast}_{in}\right],
\label{Bogoliubov-transformation}
\end{eqnarray}
where $\gamma_{n\sigma}\left(\gamma_{n\sigma}^{\dagger}\right)$ are
annihilation(creation) operators for the Bogoliubov quasiparticles in the
single-particle state $\left(n,\sigma\right)$ and satisfy canonical anticommutation
relations for fermions. The local superconducting order parameter $\Delta_{i}$ and
particle densities are obtained self-consistently via
\begin{eqnarray}
\Delta_{i} & = & -\left|U\right|\left<c_{i\uparrow}c_{i\downarrow}\right>\nonumber\\ 
          & = & -\left|U\right|\sum_{n}f\left(E_n\right)u_{in}v^{\ast}_{in}\nonumber\\
\left<n_{i\uparrow}\right> & =
&\sum_{n}f\left(E_n\right)\left|u_{n}\right|^2\nonumber\\
\left<n_{i\downarrow}\right> & = &\sum_{n}f\left(-E_n\right)\left|v_{n}\right|^2,
\label{selfconsistencyequations}
\end{eqnarray}
where $E_{n}$ is measured from the global chemical potential, $\mu$ and
$f\left(E_n\right)$ is the corresponding Fermi function. The original Hamiltonian in
Eqn.~\ref{Hamiltonian0} and the effective BdG Hamiltonian in Eqn.~\ref{BdGequation}
are manifestly invariant under $T$ and $SU(2)$. If $V = 0$ (but $V_{t} \neq 0$), the
energy eigenvalues also appear in $\pm$ pairs about the chemical potential, thus
exhibiting chiral symmetry. In the superconducting state, the disordered BdG
Hamiltonian thus belongs to the BdG chiral class.

All calculations are performed in the grand canonical ensemble scheme by keeping the
chemical potential $\mu$ constant and allowing the average electronic density $n$ to
vary freely. This approach is conceptually accurate since no adjustments are
required in order to achieve a target average density at the end of the calculation.
The disadvantage is that we need to perform calculations for a range of $\mu$ values
in order to achieve the desired $n$. However, this approach naturally provides us
results for density dependence as well. The consistency loop is started by picking a
random set for the local order parameters $\Delta_i$. Subsequently, numerical
diagonalization is carried out and the eigenvectors and eigenvalues are used to
compute the new set of $\Delta_i$, and so on until the order parameters and local
densities converge to a desired accuracy. In this work we use a convergence accuracy
of $10^{-4}$ for both the $\Delta_i$ and the charge density, i.e., we stop the
self-consistency loop when the value in $\left(k+1\right)^{{\rm{th}}}$ step differs
from the one in $k^{{\rm{th}}}$ step by less than $10^{-4}$.

\underline{Results:}

\begin{figure}[htbp]
\includegraphics[width=.90\columnwidth,angle=0,clip=true]{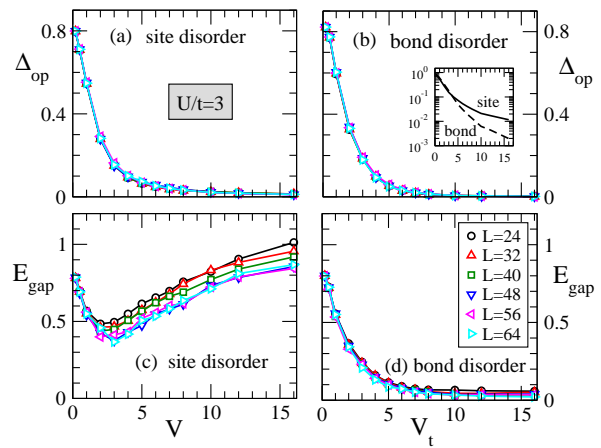}
\caption{(Color online) Dependence of the average superconducting order parameter,
(a)-(b), and the spectral gap, (c)-(d), on disorder strength for (a),(c)
diagonal(site-) disorder and (b),(d) off-diagonal(bond-) disorder for various
lattice sizes. The plots are shown for $\left|U\right|/t = 3.0$ and $n \sim 0.88$.
The disorder realization average varies from 20 copies for $L=24$ to 4 copies for
$L=64$. The inset in panel (b) compares the values of $\Delta_{{\rm{op}}}$ between
diagonal and off-diagonal disorder at large disorder bandwidths. Note that the value
for off-diagonal disorder is smaller by an order of magnitude.}
\label{fig1}
\end{figure}

We begin our discussion by showing the behavior of the spatially averaged
superconducting order parameter $\Delta_{\rm{op}}$ with disorder strength, $V$, in
Fig.~\ref{fig1}(a) and (b). Here, $\Delta_{\rm{op}}$ is defined as 
\begin{equation}
\Delta_{{\rm{op}}} = \overline{\frac{1}{N_{s}}\sum_{i}\Delta_{i}},
\label{orderparameterdefinition}
\end{equation}
where $N_{s}$ denotes the number of lattice sites and the overbar denotes an average
over different realizations of quenched disorder.

In Fig.~\ref{fig1}, panel (a) shows the behavior of a system with diagonal
(site-) disorder and panel (b) shows the behaviour of a system with
off-diagonal (bond-) disorder. In both cases, we present results for
systems where the linear lattice size takes values ranging from $L=24$ to $L=64$. It
is clear from Fig.~\ref{fig1} that $\Delta_{\rm{op}}$ has very little system-size
dependence and has, in fact, converged very well already for the system sizes $L=24$
and $L=32$. We have ascertained that, in this and all our subsequent results, all
quantities have converged by increasing the system size.


In Fig.~\ref{fig1}(a) and (b), there is already a clear qualitative difference in
the behaviour of the superconducting order parameter for the two kinds of disorder
as disorder strength is gradually increased. For the diagonal disorder case
(Fig.~\ref{fig1}(a)), the order parameter is seen to gradually converge to a
non-zero value as disorder strength $V/t$ is increased. Indeed, even when the
disorder strength is as large as the kinetic bandwidth $W_{{\rm{kin}}} = 8t$,
significant non-zero superconducting order parameter is found, when averaged over
the entire system. This was understood in terms of a phase separation of the system
between superconducting islands (with large values of $\Delta_{i}$) separated by
insulating {\textit{normal}} regions with very small $\Delta_{i}$.\cite{Ghosal:2001}

On the other hand, the figure is very different for off-diagonal disorder. As
disorder strength is increased in Fig.~\ref{fig1}(b), the order parameter supresses
rapidly. In order to clarify the difference, we plot the comparison on a logarithmic
scale in an inset in Fig.~\ref{fig1} (b). It is clear from the inset that at large
disorder strengths $\Delta_{{\rm{op}}}$ is almost an order of magnitude smaller for the
bond-disorder. In other words, at the level of mean superconducting order parameter,
bond-disorder destroys superconductivity very quickly. 
 



\begin{figure}[htbp]
\includegraphics[width=.95\columnwidth,angle=0,clip=true]{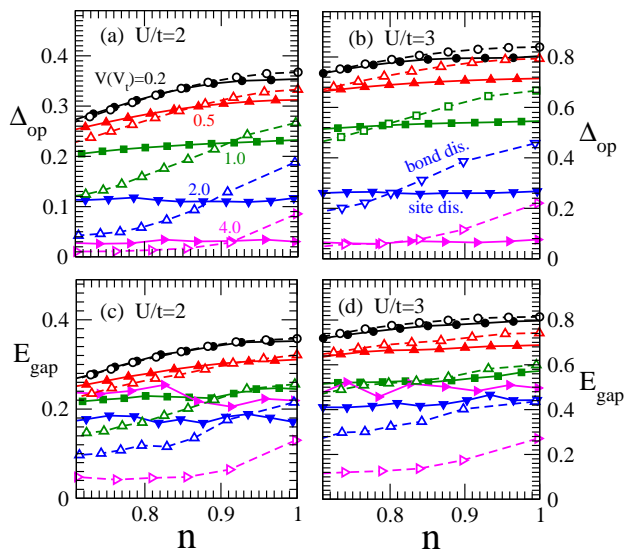}
\caption{(Color online)  Disorder and density dependence of the average
superconducting order parameter and the spectral gap for diagonal (filled symbols)
and off-diagonal (open symbols) 
disorders. Results are shown for two values of $U/t$ as indicated in the figure.
Note that the superconducting order parameter as well as the gap show a significant
density dependence in case of off-diagonal 
disorder.}
\label{fig2}
\end{figure} 

The difference in behaviour of the dirty superconductor for the two kinds of
disorder becomes even more clear if we consider how the superconducting spectral gap
responds to a change in the disorder strength. In Fig.~\ref{fig1}(c) and (d), we
plot the spectral gap $E_{{\rm{gap}}}$ defined as
\begin{equation}
E_{{\rm{gap}}} = \overline{E_{{\rm{exc}}} - E_{{\rm{g}}}}.
\label{gapdef}
\end{equation}
Here $E_{{\rm{exc}}}$ is the energy of the first excited state and $E_{{\rm{g}}}$ is
the ground state energy. The overbar, once again, denotes averaging over various
realizations of quenched disorder. Our results for the diagonal disorder (panel (c)
in Fig.~\ref{fig1}) compare very well with those presented in Ghosal {\textit{et
al}}.\cite{Ghosal:1998} This makes us confident that the behaviour of the
superconducting gap is a physical property of the system, not a numerical artefact.
Similar to Fig.~\ref{fig1}(a) and (b), we plot the spectral gap for increasing
lattice sizes. For off-diagonal disorder, Fig.~\ref{fig1}(d), the spectral gap
converges very well for all lattice sizes. Some minor size-dpendence is seen for
smaller lattices in the case of diagonal disorder, Fig.~\ref{fig1}(c), even though
the results are manifestly independent of the lattice size.

For diagonal disorder, we confirm a counter-intuitive behavior first observed by Ghosal
{\textit{et al}}.\cite{Ghosal:1998,Ghosal:2001}: the spectral gap
initially decreases with an increase in $V$, but after a minimum around $V \sim 4t$,
it monotonically increases with $V$. Ghosal
{\textit{et al}}. ascribed it to the fact that at
very high diagonal disorder, the non-interacting eigenstates are essentially
spatially local, and only those local eigenstates near the chemical potential
participate in pairing. The gap equation in this {\textit{local approximation}}
implies that the gap depends inversely on the square of the localization length
($\zeta_{\rm{loc}}$), which decreases with increasing disorder. This causes the
increase in the spectral gap with increasing disorder.

However, for off-diagonal disorder, Fig.~\ref{fig1}(d), the behaviour of the
spectral gap is strikingly different. With increasing $V_{t}$, $E_{{\rm{gap}}}$
monotonically decreases. For the largest system sizes with converged values for
$E_{{\rm{gap}}}$, it is seen to converge to a very small value - again reaffirming
the fact that off-diagonal disorder is more effective compared to diagonal disorder
in destroying superconductivity.  

\begin{figure}[htbp]
\includegraphics[width=.90\columnwidth,angle=0,clip=true]{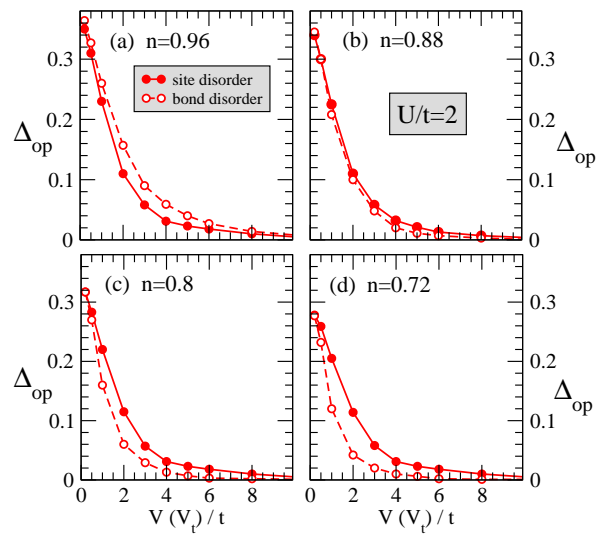}
\caption{(Color online) Disorder dependence of the superconducting order parameter
for different values of $n$, and for the diagonal and off-diagonal disorder.
Note that the relative effect of disorder depends on the average electronic density,
$n$.}
\label{fig3}
\end{figure}

In Fig.~\ref{fig2} panels (a)-(d), we introduce the dependence of the
superconducting state properties, $\Delta_{\rm{op}}$ and $E_{{\rm{gap}}}$ on the
average electron density, $n$. 
In Fig.~\ref{fig2}(a), $\Delta_{\rm{op}}$ is plotted against the electron density
for $U=2t$ and for various values of $V/t$ (full symbols with full lines)  and
$V_{t}/t$ (open symbols with dashed lines). 
We use the same set of values for both $V$ and $V_{t}$ (from $V = 0.2$ to $V =
4.0$), so that comparisons become meaningful. For $U/t = 2$,
Fig.~\ref{fig2}(a), the magnitude of $\Delta_{\rm{op}}$ 
decreases monotonically with $V/t$. At weak disorder, $\Delta_{\rm{op}}$ is weakly
density-dependent for both types of disorders. As the disorder bandwidth increases,
the $\Delta_{\rm{op}}$ for the diagonal disorder 
becomes essentially independent of $n$, whereas, the $\Delta_{\rm{op}}$ for off-diagonal
disorder continues to have a significant density dependence. 
Interestingly, there seems to be a 
crossover density $n_{cr}$ above which bond-disorder has stronger superconductivity
than site-disorder (in terms of having a larger $\Delta_{\rm{op}}$). 
For example, at half-filling, $\Delta_{\rm{op}}$ is consistently larger for $V_{t}$
than for the same $V$, for all disorder strengths investigated. Fig.~\ref{fig2}(b)
shows that the same conclusion is true 
for $U/t = 3.0$, except that $n_{cr}$ is pushed towards lower densities. However,
Fig.~\ref{fig1}(a) and (b) show that for even larger disorder strengths (for
example, at $V = V_{t} = 6t$, $n \sim 0.88$ 
and $U/t = 3.0$), this trend will again be reversed and the order parameter,
although small, will be larger for site-disorder than for bond-disorder.

In Fig.~\ref{fig2} panels (c) and (d), we plot the superconducting energy gap for
two values of $U$, $U=2.0t$ and $U=3.0t$. The non-monotonicity in the energy gap for
site-disorder is seen to occur for all 
densities and for both the values of the interaction strength, indicating that this
is a ubiquitous phenomenon for a dirty superconductor with on-site disorder. On the
other hand, it is absent for off-diagonal 
disorder for any parameter regime. The differences in the density dependence for the
two disorder types also show up in the superconducting energy gap. The plots for
diagonal disorder are essentially flat for 
$V=2t$ (filled down triangles) for both values of $U/t$. The corresponding plots for
the off-diagonal disorder show a relatively strong density dependence.

In order to emphasize the qualitative differences between the density dependence for
the two disorder types, we plot $\Delta_{\rm{op}}$ for four representative values of
the average electron density in
Fig.~\ref{fig3}, panels (a) - (d). Close to half-filling ($n=1$), the bond-disorder
has larger {\textit{average}} order parameter, while the site-disorder dominates for
all disorder strengths as we move away 
from half-filling. The difference in the density dependence points to a qualitative
difference between the manner in which the two disorder types affect the
superconducting state.  

\begin{figure}
\includegraphics[width=.95\columnwidth,angle=0,clip=true]{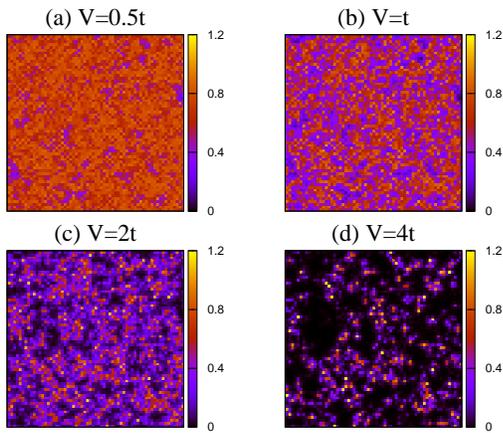}
\caption{(Color online) Real space picture of the magnitude of $\Delta_i$ for
different strengths of the diagonal disorder.}
\label{fig4}
\end{figure}
\begin{figure}[htbp]
\includegraphics[width=.95\columnwidth,angle=0,clip=true]{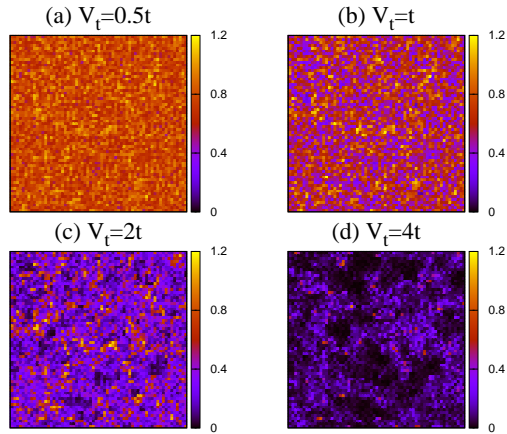}
\caption{(Color online) Real space picture of the magnitude of $\Delta_i$ for
different strengths of the off-diagonal disorder.}
\label{fig5}
\end{figure}

In order to investigate further the differences between the effects of the two
disorder types, we present the real-space data in 
Fig.~\ref{fig4} and Fig.~\ref{fig5}. We plot the real space distribution of the
local order parameter for both diagonal and off-diagonal disorder (Fig.~\ref{fig4}
and Fig.~\ref{fig5}) and at the interaction strength $U = 3.0t$ and the average
electron density $n \sim 0.88$. Looking at the real space distribution of local
quantities provide important clues about the underlying physics. In Fig.~\ref{fig4},
for diagonal disorder, we clearly see the formation of disconnected superconducting
islands which has large values of the superconducting order parameter
$\Delta_{\rm{op}}$, separated by large {\textit{normal}} regions with no
superconducting order. As the disorder strength is increased, the separation of the
homogeneous superconducting system into superconducting and normal regions become
more and more pronounced. For very large diagonal disorder, it is expected then that
the system will break up into nearly microscopic regions of superconducting islands,
interspersed by normal regions in between.  

In Fig.~\ref{fig5}, for off-diagonal disorder, we see a different behavior of the
real-space distribution of $\Delta_{\rm{op}}$ (note that we show the same disorder
bandwidths for both diagonal and off-diagonal 
disorder, so that comparisons are meaningful). For off-diagonal disorder, even for
high disorder strengths, the distribution of the superconducting order parameter is
much more homogeneous, not only in space 
but also in magnitude. There are very few lattice sites (or regions) with very high
values of $\Delta_{i}$. Thus, for the off-diagonal disorder, superconductivity is
spread more homogeneously but weakly over 
the entire system as opposed to diagonal disorder. We see a similar behavior for all
the individual disorder configurations we have investigated.

\begin{figure}[htbp]
\includegraphics[width=.95\columnwidth,angle=0,clip=true]{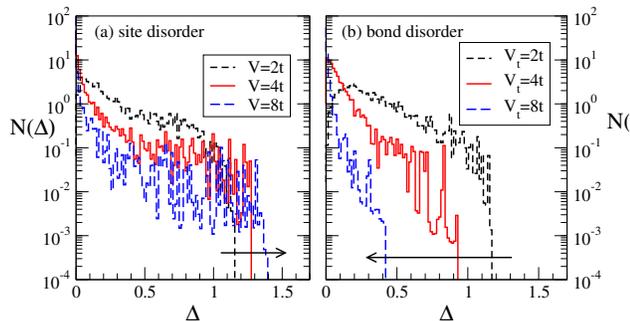}
\caption{(Color online) The distribution of local amplitudes for the superconducting
order parameters ($\Delta_i$) for (a) diagonal and (b) off-diagonal disorder. The horizontal arrows
indicate the direction in which the maximum value of $\Delta_i$ moves with increasing disorder bandwidth.}
\label{fig6}
\end{figure}

In order to make the comparison more quantitative, we plot the distribution, $N(\Delta)$, for the
local order parameters in Fig.~\ref{fig6}. $N(\Delta)$ measures the number density of sites that have the
value of superconducting order parameter belonging in the range $(\Delta-\delta/2, \Delta + \delta/2)$, with
$\delta = 0.01$.
Fig.~\ref{fig6}(a) shows the distribution for
diagonal disorder for three values of the disorder bandwidth. As the disorder
bandwidth is increasing, there are always sites having a large value for the superconducting order
parameter, contributing to the right tail of the distribution. In fact, the cut-off value for the
distribution increases, as indicated by the horizontal arrow in the figure. On the
other hand, the distribution for off-diagonal disorder shows that the
superconductivity is suppressed everywhere with increasing disorder bandwidth and no
sites remain with high values of the superconducting order parameter. Indeed, the cut-off value decreases rapidly in this case.
This behavior is consistent with the real-space pictures discussed above.

\underline{Discussion:}  In the absence of a full analytical solution to the problem
of a superconductor in the presence of strong disorder, any explanation of the
different behavior of the dirty superconductor with diagonal and off-diagonal
disorder can be at best heuristic. Typically, in a disordered system with on-site
disorder, sites with the similar disorder potential reside far away from each other.
In other words, the local potential on two neighboring sites are
{\textit{typically}} very different. This impedes the hopping of the electron over
any macroscopic distance, leading to local pair formation. As disorder bandwidth
increases, the system thus finds it advantageous to separate into smaller and
smaller microphases of superconductor and normal regions (whether the normal regions
are metallic or insulating will be discussed elsewhere). However, the same is not
true for off-diagonal disorder. A site which has a bond that impedes hopping because
of very small hopping amplitude, will typically also have another bond which leads
to a significant gain in kinetic energy, i.e., promotes hopping. Therefore, a
connected network-like region appears with an almost homogeneous order parameter
(see Fig.~\ref{fig5}(d)). Thus the pairing amplitude is uniformly spread over the
system even when the randomness in the hopping matrix element is significant. To
summarize the differences, for onsite disorder the order parameter gets
contributions from a few islands with large magnitudes of the order parameter, while
for bond disorder, the order parameter gets contributions from all over the system,
though, with increasing disorder strength, the contribution becomes uniformly weak.
This qualitative picture is very well supported by the distributions of the local
order parameter values shown in Fig.~\ref{fig6}. 

In fact, the qualitative difference
in the effective pictures also explains the difference in the density dependence
that we reported in Fig.~\ref{fig2} and Fig.~\ref{fig3}. In case of diagonal
disorder, since the microscopic regions with large value for the average
superconducting order parameter are isolated, the extra electrons find a different
island to occupy and the system remains insensitive to the electronic density. On
the other hand, bond disorder essentially leads to a connected sub-system which can
be considered as a sub-system of a new shape and size with weak disorder and therefore has a density
dependence similar to that in a weakly disordered system. We propose qualitatively
distinct pictures of how disorder destoys superconductivity in a site-
{\textit{vs.}} a bond-disordered system. We believe that this also has implications
for how superconductivity will be destroyed at the critical point between the
superconductor and the Anderson insulator. In agreement with previous theoretical
work,\cite{Grinstein:1990, Ghosal:1998, Ghosal:2001}, for the diagonal disorder, the
non-vanishing spectral gap implies that even with increasing disorder, low-energy
{\textit{fermionic}} quasiparticles are not generated. Thus the superconductivity
will be destroyed by increasing phase fluctuations among the superconducting
islands. The critical theory will be {\textit{bosonic}}, and in $d=2$, will consist
of a diffusing metal of Cooper pairs at the quantum critical point itself. On the
other hand, the superconductor in the presence of strong off-diagonal disorder will
be destroyed in a different fashion, with the order parameter uniformly vanishing
across the system, and with the presence of fermionic quasiparticles at the
transition (for a recent experimental work on the possibility of the existence of a
bosonic and a fermionic SIT, see Hollen {\textit{et al}}).\cite{Hollen:2013}
Unfortunately, in conventional solid state systems, the tuning of the type of
disorder is experimentally very difficult. However, with the advent of the cold atom
systems, we hope different scenarios with different types of disorder can be tested
experimentally.   

We acknowledge many useful discussions with M. Jiang, R. T. Scalettar, K. Byczuk and
D. Vollhardt. This research was partially supported by the Deutsche
Forschungsgemeinschaft through the TRR 80 (PBC), the Chennai center of the Indian
Statistical Institute (PBC) and the Department of Science and Technology, India
(SK).            





\end{document}